\documentclass[seceq,preprint]{ptptex}

\usepackage{graphicx}
\usepackage{wrapft}

\preprintnumber[3cm]{SMUHEP/04-01}

\title{ Extended Hamiltonian Formalism of the \\
Pure Space-Like Axial Gauge Schwinger Model II}
\subtitle{}    

\author{ Yuji {\sc Nakawaki} and Gary {\sc McCartor}$^*$
}

\inst{Division of Physics and Mathematics, Faculty of Engineering \\
 Setsunan University, Osaka 572-8508, Japan
\\
$^*$Department of Physics, Southern Methodist University  \\
Dallas TX 75275,  USA}

\recdate{
}

\abst{ Canonical methods are not sufficient to properly quantize space-like 
axial gauges. In this paper, we obtain guiding principles which allow the 
construction of an extended Hamiltonian formalism for pure space-like axial 
gauge fields.  To do so, we clarify the general role residual gauge fields 
play in the space-like axial gauge Schwinger model. In all the calculations we 
fix the gauge using a rule, $n{\cdot}A=0$, where $n$ is a space-like constant 
vector and we refer to its direction as $x_-$. Then, to begin with, we 
construct a formulation in which the quantization surface is space-like but 
not parallel to the direction of $n$. The quantization surface has a parameter 
which allows us to rotate it, but when we do so we keep the direction of the 
gauge field fixed. In that formulation we can use canonical methods. We 
bosonize the model to simplify the investigation. We find that the 
antiderivative, $({\partial}_-)^{-1}$, is ill-defined whatever quantization 
coordinates we use as long as the direction of $n$ is space-like. We find that 
the physical part of the dipole ghost field includes infrared divergences. 
However, we also find that if we introduce residual gauge fields in such a way 
that the dipole ghost field satisfies the canonical commutation relations, 
then the residual gauge fields are determined so as to regularize the infrared 
divergences contained in the physical part.  The propagators then take the 
form prescribed by Mandelstam and Leibbrandt.  We make use of these properties 
to develop guiding principles which allow us to construct consistent operator 
solutions in the pure space-like case where the quantization surface is 
parallel to the direction of $n$ and canonical methods do not suffice.  }   

\begin{document}

\maketitle

\section{Introduction}

In a previous paper, which is hereafter referred to as I\cite{rf:1}, we 
constructed an extended Hamiltonian formalism with which we found a family of 
solutions to the Schwinger model. The solutions were of the axial or temporal 
gauge type. To consider the problem generally, we specified the gauge fixing 
direction by the constant vector $n^{\mu}=(n^0,n^3)=(\cos \theta,\sin \theta)$.  At the same time we introduced $+-$-coordinates $x^{\mu}=(x^+,x^-)$, where
\begin{equation}
 x^+=x^0{\sin}{\theta}+x^3{\cos}{\theta}, \quad
x^-=x^0{\cos}{\theta}-x^3{\sin}{\theta}  \label{eq:1.1}
\end{equation}
With those definitions, the gauge fixing condition
\begin{equation}
 A_-=n{\cdot}A=A_0\cos \theta -A_3\sin \theta=0  \label{eq:1.2}
\end{equation}
is that of an axial or temporal gauge.  In our formulation, the temporal 
and axial gauges in ordinary coordinates correspond, respectively, to 
$\theta=0$ and  $\theta
=\frac{\pi}{2}$, while the light-front formulation corresponds to
 $\theta =\frac{\pi}{4}$.  We found that in the region 
$0{\leqq}{\theta}<\frac{\pi}{4},\;x^-$ should be taken as the evolution 
parameter and we constructed the canonical temporal gauge solutions.  In that 
case, we found that there exist residual gauge fields which depend only on 
$x^+$.  These residual gauge fields are therefore static canonical variables.  
By continuation, we  obtained an operator solution in the axial region, 
$\frac{\pi}{4}<{\theta}<\frac{\pi}{2}$, where $x^+$ should be taken as 
the evolution parameter. In that case, we find that there are infrared 
divergences associated with the physical degrees of freedom.   These infrared 
divergences are regularized by the residual gauge fields. Among other results, 
we found that the Hamiltonian for the residual gauge fields must be calculated 
by integrating the divergence equation of the energy-momentum tensor over a 
suitable closed surface. Because the residual gauge fields do not depend on 
the initial value surface, $x^-$, the ($x^-\to{\pm}\infty)$ contributions from 
these fields have to be kept.\cite{rf:2} In that way, we obtained the 
Hamiltonian, which includes a part from integrating a density involving the 
residual gauge fields over $x^-=$ constant.   

In I, we found the solutions in the axial gauge region only by continuation 
from the temporal gauge region. In this paper we consider the problem of 
finding the axial gauge solutions directly; by quantizing on the surface 
$x^+ = 0$. This axial gauge formulation involves constrained fields and 
traditionally these constrained fields are eliminated in terms of physical 
degrees of freedom. That elimination requires that we introduce 
antiderivatives which can introduce infrared divergences.\cite{rf:3} In spite 
of extensive studies\cite{rf:4}, overcoming the infrared difficulties has 
remained as an open issue. In the present work we find that the residual gauge 
fields are essential to controlling  the infrared divergences. These fields 
may be viewed as integration constants associated with solving the constraint 
equations and they are necessary to give the correct prescription for the 
required antiderivatives. Quantizing the residual gauge fields is itself an 
interesting subject. This is because they depend on the evolution parameter in 
such a way that they cannot be canonical variables. A first step in this 
direction was made by McCartor and Robertson\cite{rf:5} in the light-front 
formulation of free abelian gauge fields.

To begin with, we consider a generalization of the models considered in I. We 
shall keep the constant vector at the fixed space-like direction and take the 
quantization surface to be space like, but will not have the constant vector 
lie parallel to the quantization surface. In that framework we can implement 
the canonical procedure. We then use the operator solutions found in such 
cases to clarify the dependence of the operator solutions on the quantization 
coordinates. We find that there are the residual gauge fields allowed by the 
fixed gauge choice and we can also use the operator solutions to clarify the 
general roles the residual gauge fields play in these axial gauge solutions. 
To implement these ideas we introduce another set of coordinates, $x^{\mu}=
(x^{\tau},x^{\sigma})$, defined by 
\begin{equation}
 x^{\tau}=x^0{\sin}{\phi}+x^3{\cos}{\phi}, \quad
x^{\sigma}=x^0{\cos}{\phi}-x^3{\sin}{\phi}.   \label{eq:1.3}
\end{equation}
In these coordinates the gauge fixing condition and the constant 
vector are expressed, respectively, as 
\begin{eqnarray}
&A_-=\sin(\phi-\theta)A_{\tau}+\cos(\phi-\theta)A_{\sigma}=0,&
\label{eq:1.4} \\  
&n^{\mu}=(n^{\tau},n^{\sigma})=(\sin(\phi-\theta),\cos(\phi-\theta)).& 
\end{eqnarray}
To simplify our investigation, we bosonize the Schwinger model and avoid 
quantizing the coupled system of fermi fields and gauge fields.  The 
solutions contain a dipole ghost field, $X$, which contains both physical 
fields and residual gauge fields. In the space-like formulations where the 
constant vector is not parallel to the quantization surface we can employ 
$A_{\sigma}$ and the dipole ghost field $X$ as canonical variables and 
construct a canonical formulation without encountering any of the difficulties 
inherent in the pure space-like (PSL) axial gauge formulations where the 
constant vector is proportional to the quantization surface. We show that the 
physical part of $X$ is uniquely determined by the gauge choice, while the 
residual gauge part, which reveals manifest quantization coordinate 
dependence, is determined by requiring that $X$ satisfy the canonical 
commutation conditions.  It turns out that $(n{\cdot}{\partial})^{-1}=
({\partial}_-)^{-1}$ is ill-defined irrespective of the quantization 
coordinates as long as the gauge fixing direction is space-like ($n^2<0$). It 
follow from this that the physical part of $X$ gives rise to infrared 
divergences irrespective of the quantization coordinates  as long as the gauge 
fixing direction is space-like.  However, if we introduce the residual gauge 
fields in such a way that $X$ satisfies the canonical commutation conditions, 
then the residual gauge part  automatically regularizes the infrared 
divergences resulting from the physical part.  As a consequence, the 
$x^{\tau}$-time ordered propagator for $X$ takes the form prescribed by 
Mandelstam\cite{rf:6} and by Leibbrandt\cite{rf:7} (ML prescription). In this 
way we see that the residual gauge degrees of freedom are indispensable to 
formulate the space-like axial gauge Schwinger model in a way that is free 
from infrared divergences.  

We remark here that canonical formulations in ordinary coordinates were 
constructed for the case $n^2=0$ by Bassetto et al\cite{rf:8} and for 
the case $n^0{\ne}0,\;n^2<0$ by Lazzizzera\cite{rf:9}.  These authors showed 
that to implement the ML prescription and to regularize the infrared 
divergences, residual gauge fields are indispensable.   

Having found the solutions to the axial gauge formulations in the cases where
the constant vector is not parallel to the the quantization surface, we 
turn to the the pure space-like case. The PSL case cannot be reached by taking 
the limit $\phi{\to}\theta$.  This reflects the fact that we cannot construct 
the canonical formulation in the PSL case because residual gauge fields cannot 
be  canonical fields; only $X$ and its conjugate remain as unconstrained 
canonical fields.  We circumvent this difficulty by using the properties 
of the dipole ghost fields found in \S~2 as guiding principles.  We show that 
operator solutions can be constructed by following these guiding principles.  
When these operator solutions are constructed they agree with ones given in I.

The paper is organized as follows: In ${\S}~2$, we bosonize the space-like 
axial gauge Schwinger model and construct the canonical formulation in 
$\tau\sigma$-coordinates.  In ${\S}~3$ we show that our canonical formulation 
is free from infrared difficulties. In ${\S}~4$, we carry out the quantization 
of the PSL case and construct the solution. Section 5 is devoted to concluding 
remarks.

In this paper we keep $\phi$ in the axial region
 $\frac{\pi}{4}<\phi{\leqq}\frac{\pi}{2}$ and use the following conventions in 
$\tau\sigma$-coordinates: 
$$g_{\tau\tau}=-\cos 2{\phi}, \quad  g_{\sigma\tau}=g_{\tau\sigma}
=\sin 2{\phi}, \quad g_{\sigma\sigma}=\cos 2{\phi},  $$
$$g^{\tau\tau}=-\cos 2{\phi}, \quad  g^{\sigma\tau}=g^{\tau\sigma}
=\sin 2{\phi}, \quad g^{\sigma\sigma}=\cos 2{\phi},  $$
$$ {\gamma}^0={\sigma}_1, \;{\gamma}^3=i{\sigma}_2, \;
 {\gamma}^5=-{\sigma}_3, 
$$
$${\gamma}^{\tau}={\gamma}^0\sin {\phi}+{\gamma}^3\cos {\phi},\quad
{\gamma}^{\sigma}={\gamma}^0\cos {\phi}-{\gamma}^3\sin {\phi}.$$

\section{Equivalent bosonization of space-like axial gauge Schwinger model}
\subsection{Field equation of the dipole ghost field}
The space-like axial gauge Schwinger model is defined  by the Lagrangian 
\begin{equation}
L=-\frac{1}{4}F_{{\mu}{\nu}}F^{{\mu}{\nu}}-B(n{\cdot}A) 
+i\bar{\psi}{\gamma}^{\mu}({\partial}_{\mu}+ieA_{\mu}){\psi}  
\label{eq:2.1}
\end{equation} 
where $B$ is the Nakanishi-Lautrup field in noncovariant formulations.
\cite{rf:10} From the Lagrangian we derive the field equations
\begin{equation}
{\partial}_{\mu}F^{{\mu}{\nu}}=n^{\nu}B +J^{\nu}, \quad 
J^{\nu}=e\bar{\psi}{\gamma}^{\nu}{\psi} \label{eq:2.2}
\end{equation}
\begin{equation}
i{\gamma}^{\mu}({\partial}_{\mu}+ieA_{\mu}){\psi}=0, \label{eq:2.3}
\end{equation}
and the gauge fixing condition (\ref{eq:1.4}).  The field
 equation 
of $B$,
\begin{equation}
(n{\cdot}{\partial})B={\partial}_-B=(\sin(\phi-\theta){\partial}_{\tau}
+\cos(\phi-\theta){\partial}_{\sigma})B=0, \label{eq:2.4}
\end{equation}
follows from operating on (\ref{eq:2.2}) with ${\partial}_{\nu}$.

Let's first obtain the field equation of the dipole ghost field $X$.  
We now know that consistent operator solutions of Schwinger
 model\cite{rf:11} 
can be constructed by regularizing the vector current by means of the gauge 
invariant point-splitting procedure\cite{rf:12}. We will therefore regularize 
$J^{\mu}$ in the same manner in the present paper.  With that regularization, 
the vector current is given by  
\begin{equation}
J^{\mu}=j^{\mu}-m^2A^{\mu}  \label{eq:2.5}
\end{equation}
where $m^2=\frac{e^2}{\pi}$ and $j^{\mu}$ is the part given as the bilinear 
product of the ${\psi}$.
We now observe  that Eq.(\ref{eq:2.3}) is massless; that is, $j^{\mu}$ 
satisfies ${\varepsilon}^{{\mu}{\nu}}{\partial}_{\mu}j_{\nu}=0$.  Therefore 
we can define $j^{\mu}$ as the gradient of the dipole ghost field $X$:   
\begin{equation}
j_{\mu}=m{\partial}_{\mu}X.  \label{eq:2.6}
\end{equation}
Substituting (\ref{eq:2.6}) into (\ref{eq:2.5}) and then using current 
conservation, ${\partial}_{\mu}J^{\mu}=0$, we obtain   
\begin{equation}
m{\square}X=m^2{\partial}^{\mu}A_{\mu}.  \label{eq:2.7}
\end{equation}
Substituting (\ref{eq:2.5}), (\ref{eq:2.6}) and (\ref{eq:2.7}) into 
(\ref{eq:2.2}) we get 
\begin{equation}
({\square}+m^2)(A^{\nu}-\frac{1}{m}{\partial}^{\nu}X)=n^{\nu}B.
 \label{eq:2.8}
\end{equation}
Finally, operating with $n_{\nu}$ on (\ref{eq:2.8}) and using $n{\cdot}A=0$ we 
derive the field equation of the dipole ghost field $X$
\begin{equation}
({\square}+m^2)({\partial}_-X)=-mn^2B.  \label{eq:2.9}
\end{equation}

\subsection{Bosonization of the generalized axial gauge Schwinger model}

Now we can employ Eqs.(\ref{eq:2.7}) and (\ref{eq:2.8}) as a guiding 
principles to obtain the Lagrangian for the equivalent bosonized model.  
These equations are derived from
\begin{equation}
L=-\frac{1}{4}F_{{\mu}{\nu}}F^{{\mu}{\nu}}-B(n{\cdot}A) 
+\frac{1}{2}{\partial}_{\mu}X{\partial}^{\mu}X-m{\partial}_{\mu}XA^{\mu}
+\frac{m^2}{2}A_{\mu}A^{\mu}, \label{eq:2.10}
\end{equation} 
which justifies the use of (\ref{eq:2.10}) as the Lagrangian in the present 
variables.  From this Lagrangian, we see that in the axial region, $\pi/4<\phi
{\leqq}\pi/2$, where $x^{\tau}$ is chosen as the evolution parameter, the
fundamental fields are $A_{\sigma}$ and $X$.   $A_{\tau}$ 
is a dependent field as long as $\phi{\ne}\theta$.  

Canonical conjugate momenta are found from the Lagrangian to be
\begin{equation}
 {\pi}^{\tau}=0, \;\; {\pi}^{\sigma}=F_{\tau\sigma},\;\;
 {\pi}_{\scriptscriptstyle B}=0,\;\;{\pi}_{\scriptscriptstyle X}
={\partial}^{\tau}X-mA^{\tau}.  \label{eq:2.11}
\end{equation}
Therefore we can choose $A_{\sigma},\;X,\;{\pi}^{\sigma}$ and 
${\pi}_{\scriptscriptstyle X}$ as independent canonical variables and 
express the dependent degrees of freedom as  
\begin{equation}
A_{\tau}=\cot ({\theta}-{\phi})A_{\sigma},\;
B=({\partial}_{\sigma}{\pi}^{\sigma}-m{\pi}_{\scriptscriptstyle
 X})/n^{\tau}.  
\label{eq:2.12}
\end{equation}
Consequently, equal $x^{\tau}$-time canonical quantization conditions can be 
imposed on the independent canonical variables; the nonvanishing commutators 
are
\begin{equation}
[A_{\sigma}(x),\pi^{\sigma}(y)]=i\delta(x^{\sigma}-y^{\sigma}),\;\;
[X(x),\pi_{\scriptscriptstyle X}(y)]=i\delta(x^{\sigma}-y^{\sigma}).  
\label{eq:2.13}
\end{equation}
For later convenience we give here the equal $x^{\tau}$-time commutation
 relations 
of $B$:
\begin{eqnarray}
&[{\pi}^{\sigma}(x),B(y)]=[{\pi}_{\scriptscriptstyle X}(x),B(y)]
=[B(x),B(y)]=0,& \nonumber \\
&[X(x),B(y)]=-i\frac{m}{n^{\tau}}\delta(x^{\sigma}-y^{\sigma}),\;\;
[A_{\sigma}(x),B(y)]=-\frac{i}{n^{\tau}}{\partial}_{\sigma}
\delta(x^{\sigma}-y^{\sigma}).& \label{eq:2.14}
\end{eqnarray}

\subsection{Expression of the dipole ghost field}

 Now that the canonical formulation is given, we proceed to solving 
Eq.(\ref{eq:2.8}).  To obtain a particular solution, we make use of the 
fact that, due to (\ref{eq:2.4}), $B$ satisfies
\begin{equation}
({\square}+m^2)B=\left(m^2-\frac{n^2{\partial}_{\sigma}^{\;2}}
{\sin^2(\phi-\theta)}\right)B=(m^2-n^2{\partial}_+^{\;2})B.  \label{eq:2.15}
 \end{equation}
Here and in what follows, we denotes, for brevity, $-\frac{{\partial}_{\sigma}}
{\sin(\phi-\theta)}$ as ${\partial}_+$ when it is applied to operators 
dependent on only $x^+$.  It follows immediately that 
a particular solution to equation (\ref{eq:2.8}), for the quantity $A^{\nu}-
\frac{1}{m}{\partial}^{\nu}X$, is
$\frac{n^{\mu}}{m^2-n^2{\partial}_+^{\;2}}B.$

To specify the remaining homogeneous part, which satisfies the free 
D'Alembert's equation of mass $m$, we take account of the fact 
that $F_{\tau\sigma}$ is gauge invariant and satisfies $F_{\tau\sigma}
=F_{+-}$.  We see from this that $F_{\tau\sigma}$ is independent of the 
quantization coordinates and therefore agrees with one given by the 
solution in I
\begin{equation}
F_{\tau\sigma}=
m\tilde{\Sigma}+\frac{n^2}{m^2-n^2{\partial}_+^{\;2}}
{\partial}_+B \label{eq:2.16}
\end{equation}
where $\tilde{\Sigma}$ is the Schwinger field of mass $m$.  
We can easily see that (\ref{eq:2.16}) can be derived from the following 
expression for $A^{\nu}-\frac{1}{m}{\partial}^{\nu}X$:
\begin{equation}
A^{\mu}-\frac{1}{m}{\partial}^{\mu}X=\frac{n^{\mu}}{m^2-n^2{\partial}_+^{\;2}}B
-{\varepsilon}^{\mu\nu}\frac{{\partial}_{\nu}\tilde{\Sigma}}{m}.\label{eq:2.17}
\end{equation}
where  ${\varepsilon}^{\tau\sigma}=-{\varepsilon}^{\sigma\tau}=1,
{\varepsilon}^{\tau\tau}={\varepsilon}^{\sigma\sigma}=0.$

It is useful to point out here that the right hand side of (\ref{eq:2.17})
 can be written in the following, divergence free, form
\begin{equation}
\frac{n^{\mu}}{m^2-n^2{\partial}_+^{\;2}}B
-{\varepsilon}^{\mu\nu}\frac{{\partial}_{\nu}\tilde{\Sigma}}{m}
=-\frac{1}{m}{\varepsilon}^{\mu\nu}{\partial}_{\nu}{\lambda} \label{eq:2.18}
\end{equation}
where 
\begin{equation}
\lambda=\tilde{\Sigma}-\frac{mn^{\tau}}{m^2-n^2{\partial}_+^{\;2}}
{\partial}_{\sigma}^{-1}B
\label{eq:2.19}
\end{equation}
and the operator ${\partial}_{\sigma}^{-1}$ is defined by
\begin{equation}
({\partial}_{\sigma})^{-1}f(x)=\frac{1}{2}\int_{-{\infty}}^{\infty}
dy^{\sigma}{\varepsilon}(x^{\sigma}-y^{\sigma})f(x^{\tau},y^{\sigma})  
\label{eq:2.20}
\end{equation}
which imposes, in effect, the principal value regularization.
It follows from (\ref{eq:2.5}), (\ref{eq:2.6}), (\ref{eq:2.17}) and 
(\ref{eq:2.18}) that 
$A^{\mu}$ and $J^{\mu}$ can be written as 
\begin{equation}
A^{\mu}=\frac{1}{m}({\partial}^{\mu}X-{\varepsilon}^{\mu\nu}{\partial}_{\nu}
{\lambda}),\;\;J^{\mu}=m{\varepsilon}^{\mu\nu}{\partial}_{\nu}{\lambda}.
\label{eq:2.21}
\end{equation}
We can now verify that $\tilde{\Sigma}$ and 
${\partial}^{\tau}\tilde{\Sigma}$  
satisfy canonical equal $x^{\tau}$-time commutation relations 
\begin{eqnarray}
&[\tilde{\Sigma}(x),\tilde{\Sigma}(y)]=[{\partial}^{\tau}\tilde{\Sigma}(x),
{\partial}^{\tau}\tilde{\Sigma}(y)]=0,\;\;[\tilde{\Sigma}(x),
{\partial}^{\tau}\tilde{\Sigma}(y)]=i{\delta}(x^{\sigma}-y^{\sigma}),& \\
&[B(x),\tilde{\Sigma}(y)]=[B(x),{\partial}^{\tau}\tilde{\Sigma}(y)]=0&
\label{eq:2.23}
\end{eqnarray}
by using their expressions in terms of the canonical variables: 
\begin{equation}
\tilde{\Sigma}=\frac{1}{m}({\pi}^{\sigma}-\frac{n^2}{m^2-n^2{\partial}_+^{\;2}}
{\partial}_+B),\;\;
{\partial}^{\tau}\tilde{\Sigma}={\partial}_{\sigma}X-mA_{\sigma}
+\frac{mn_{\sigma}}{m^2-n^2{\partial}_+^{\;2}}B.  \label{eq:2.24}
\end{equation}

Let's next obtain an expression for $X$.  To this aim we multiply 
(\ref{eq:2.17}) by 
$n_{\mu}$ and use $n{\cdot}A=A_-=0$ and $n{\cdot}{\partial}={\partial}_-$.  
We then get
\begin{equation}
{\partial}_-X=-\frac{mn^2}{m^2-n^2{\partial}_+^{\;2}}B
+{\varepsilon}^{\mu\nu}n_{\mu}{\partial}_{\nu}\tilde{\Sigma}
=-\frac{mn^2}{m^2-n^2{\partial}_+^{\;2}}B+{\partial}^+\tilde{\Sigma}
\label{eq:2.25}
\end{equation}
and see that $X$ is obtained by integrating (\ref{eq:2.25}) with respect 
to $x^-$.   The first term has to be carefully integrated.  At first sight it 
seems that a linear function of $x^-$ is included because the first term 
depends on only $x^+$.  However it turns out that if $X$ has such  term, then 
the equal $x^{\tau}$-time canonical commutation relations of $X$ 
are not satisfied.  We use the possibility of adding arbitrary functions of $x^+$ to write the integral of the first term as
$-\frac{x^{\tau}}{n^{\tau}}\frac{mn^2}{m^2-n^2{\partial}_+^{\;2}}B.$
To integrate the second term, we make use of the antiderivative 
$({\partial}_-)^{-1}$ defined by  
\begin{equation}
\frac{1}{{\partial}_-}\tilde{\Sigma}=-\frac{n^{\tau}{\partial}^{\tau}+
n_{\sigma}{\partial}_{\sigma}}{m^2\sin^2(\phi-\theta)
-n^2{\partial}_{\sigma}^{\;2}}\tilde{\Sigma}.  \label{eq:2.26}
\end{equation} 
We can show that (\ref{eq:2.26}) is correct by operating on both sides with 
$ {\partial}_-=n^{\tau}{\partial}_{\tau}+n^{\sigma}{\partial}_{\sigma}
=\frac{n^{\tau}{\partial}^{\tau}-n_{\sigma}{\partial}_{\sigma}}{-\cos2\phi}$ 
and using mass shell condition 
$\{({\partial}^{\tau})^2-{\partial}_{\sigma}^{\;2}-\cos2\phi m^2\}
\tilde{\Sigma}=0$.  
We thus obtain the general solution which we write in the form
$$
X=-\frac{x^{\tau}}{n^{\tau}}\frac{mn^2}{m^2-n^2{\partial}_+^{\;2}}B
+\frac{{\partial}^+}{{\partial}_-}\tilde{\Sigma}
+{\rm integration\;\;constant}.  $$

The integration constant is determined in the following way: To obtain 
the first commutation relation in the second line of (\ref{eq:2.14}), we need 
an operator which does not commute with $B$; that is because $B$ commutes with 
both $\tilde{\Sigma}$ and ${\partial}^{\tau}\tilde{\Sigma}$ as seen in 
(\ref{eq:2.23}) and so with $\frac{{\partial}^+}{{\partial}_-}\tilde{\Sigma}$, 
which is described as
\begin{equation}
\frac{{\partial}^+}{{\partial}_-}\tilde{\Sigma}=
-\frac{m^2n^{\tau}n_{\sigma}+n^2{\partial}^{\tau}{\partial}_{\sigma}}
{m^2\sin^2(\phi-\theta)-n^2{\partial}_{\sigma}^{\;2}}\tilde{\Sigma}.
\label{eq:2.27}
\end{equation} 
Therefore we must introduce another field, $C$, which depends on only 
$x^+$.  To obtain the relation $[X(x),X(y)] = 0$ when $x^{\tau}=y^{\tau}$, we 
need an extra term.  That is because it is natural to assume that $C$ commutes 
with $\tilde{\Sigma}$ and ${\partial}^{\tau}\tilde{\Sigma}$, and 
because the commutator $[\frac{{\partial}^+}{{\partial}_-}\tilde{\Sigma}(x),
\frac{{\partial}^+}{{\partial}_-}\tilde{\Sigma}(y)]$ does not vanish
when $x^{\tau}=y^{\tau}$. We find that if we parameterize the integration 
constant in the form
\begin{equation}
X=\frac{{\partial}^+}{{\partial}_-}\tilde{\Sigma}
+\frac{m}{m^2-n^2{\partial}_+^{\;2}}
\left(C-\frac{n^2x^{\tau}}{n^{\tau}}B+\frac{n^2n_{\sigma}}
{m^2\sin^2(\phi-\theta)-n^2{\partial}_{\sigma}^{\;2}}{\partial}_{\sigma}B,
\right)   \label{eq:2.28}
\end{equation}
then the canonical commutation conditions yield the following equal 
$x^{\tau}$-time commutation relations for $C$:
\begin{eqnarray}
&[C(x),C(y)]=0,\;\;[B(x),\frac{1}{m^2-n^2{\partial}_+^{\;2}}C(y)]
=i\frac{1}{n^{\tau}}{\delta}(x^{\sigma}-y^{\sigma}),& 
\nonumber \\
&[C(x),\tilde{\Sigma}(y)]=[C(x),{\partial}^{\tau}\tilde{\Sigma}(y)]=0.& 
\label{eq:2.29}
\end{eqnarray}
Substituting (\ref{eq:2.28}) into (\ref{eq:2.17}) then yields an explicit 
expression for $A_{\mu}$:
\begin{eqnarray}
A_{\mu}&=&{\varepsilon}_{\mu\nu}n^{\nu}\frac{m}
{{\partial}_-}\tilde{\Sigma}
+\frac{n_{\mu}}{m^2-n^2{\partial}_+^{\;2}}B \nonumber \\
\hspace*{-0.5mm}&+&\hspace*{-0.5mm}\frac{{\partial}_{\mu}}
{m^2-n^2{\partial}_+^{\;2}}
\left(C-\frac{n^2x^{\tau}}{n^{\tau}}B+\frac{n^2n_{\sigma}}
{m^2\sin^2(\phi-\theta)-n^2{\partial}_{\sigma}^{\;2}}{\partial}_{\sigma}B
\right) .   \label{eq:2.30}
\end{eqnarray}

In this way we see that the residual gauge fields are 
indispensable to preserve the canonical commutation relations and that the 
residual gauge part of $X$ must include an explicit dependence on the
quantization coordinates.  
We close this subsection by pointing out how infrared divergences appear 
in our formulation.  As is seen from (\ref{eq:2.30}), 
the inverse of the operator $m^2\sin^2(\phi-\theta)- n^2
{\partial}_{\sigma}^{\;2}$ is applied to both $\tilde{\Sigma} $ and to the 
residual gauge fields.  This inverse operator gives rise to infrared 
divergences because $n^2=\cos 2\theta<0$ in the range  $\frac{\pi}{4}
<\theta<\frac{\pi}{2}$.  So that operator becomes singular in our range. We 
show in next section that the infrared divergences resulting from the physical 
field are cancelled by infrared divergences from the residual gauge part.

\subsection{Fermion field operator}
Now that we have the explicit expression for $A_{\mu}$, we can construct the 
fermion field operators in the same way as in I.  From the expression for 
$A^{\mu}$ in (\ref{eq:2.21}), we see that the fermion operators are formally 
given by 
$${\psi}_{\alpha}(x)=\frac{Z_{\alpha}}{\sqrt{({\gamma}^0{\gamma}^{\tau})_
{{\alpha}{\alpha}}}}
\exp[-i\sqrt{\pi}{\Lambda}_{\alpha}(x)], \;\; (\alpha=1,2)  $$
where $Z_{\alpha}$ is normalization constant and   
\begin{equation}
{\Lambda}_{\alpha}(x)=X(x)+(-1)^{\alpha}{\lambda}(x).  \label{eq:2.31}
\end{equation}
We need to rewrite the formal solution into a normal ordered product.  
However, if we simply normal order the exponential and then calculate the 
canonical anticommutation relations, we find another infrared  divergence 
inherent in two-dimensional massless scalar fields.  In our formulation it 
results from the singular operator, ${\partial}^{-1}_{\sigma}B$, in $\lambda$ 
in (\ref{eq:2.19}).  We overcome this difficulty  by not rewriting the 
infrared parts of the singular operator and its conjugate operator into normal 
ordered form.\cite{rf:13} In what follows, we keep $\phi>\theta$ and, to 
incorporate the ML prescription, we employ the following representations of 
$B$ and $C$:  
\begin{eqnarray}
&B(x)=\frac{m}{n^{\tau}\sqrt{2\pi}}\int^{\infty}_{-\infty}dk_{\sigma}
\theta(-k_{\sigma})\sqrt{|k_{\sigma}|}
\{ B(k_{\sigma})e^{-ik{\cdot}x} 
+ B^{\ast}(k_{\sigma})e^{ik{\cdot}x} \},&\nonumber \\ 
&\frac{m}{m^2-n^2{\partial}_+^{\;2}}C(x)\hspace*{-1mm}=
\hspace*{-1mm}\frac{i}{\sqrt{2\pi}}\int^{\infty}_{-\infty} 
\frac{dk_{\sigma}}{\sqrt{|k_{\sigma}|}}
\theta(-k_{\sigma})\{ C(k_{\sigma})
e^{-ik{\cdot}x} - C^{\ast}(k_{\sigma})e^{ik{\cdot}x} \},& \label{eq:2.32}
\end{eqnarray}
where $k_{\tau}=\cot(\theta-\phi)k_{\sigma}$, creation and annihilation 
operators satisfy
\begin{equation}
[B(k_{\sigma}),C^{\ast}(q_{\sigma})]=[C(k_{\sigma}),B^{\ast}(q_{\sigma})]=
-{\delta}(k_{\sigma}-q_{\sigma}),  \label{eq:2.33}
\end{equation}
and all other commutators are zero.  
These relations allow us to define the physical subspace, $V$, by
\begin{equation}
V=\{\;\;|{\rm phys}{\rangle}\;\;|\;\; B(k_{\sigma})|{\rm phys}{\rangle}
=0\;\; \}.  \label{eq:2.34}
\end{equation}
and to define the infrared part, ${\Lambda}_{\alpha}^{(0)}$, of 
${\Lambda}_{\alpha}$  by
\begin{equation}
{\Lambda}_{\alpha}^{(0)}=\frac{i}{\sqrt{2{\pi}}}\int^0_{-\kappa}
\frac{dk_{\sigma}}{\sqrt{|k_{\sigma}|}}
\{C(k_{\sigma})-C^{\ast}(k_{\sigma})+(-1)^{\alpha}(B(k_{\sigma})
-B^{\ast}(k_{\sigma}))\} \label{eq:2.35}
\end{equation}
where ${\kappa}$ is a small positive constant.  

Now we can define the fermion field operators to be
\begin{equation}
{\psi}_{\alpha}(x)=\frac{Z_{\alpha}}{\sqrt{({\gamma}^0{\gamma}^{\tau})_
{{\alpha}{\alpha}}}}
{\rm exp}[-i\sqrt{\pi}{\Lambda}_{{\alpha}r}^{(-)}(x)]{\sigma}_{\alpha}
{\rm exp}[-i\sqrt{\pi}{\Lambda}_{{\alpha}r}^{(+)}(x)]  \label{eq:2.36}
\end{equation}
where ${\Lambda}_{{\alpha}r}^{(-)}$ and ${\Lambda}_{{\alpha}r}^{(+)}$  
are creation and annihilation operator parts of ${\Lambda}_{{\alpha}r}
{\equiv}{\Lambda}_{\alpha}-{\Lambda}_{\alpha}^{(0)}$ and
\begin{equation}
{\sigma}_{\alpha}=\exp\left[-i\sqrt{\pi}\left({\Lambda}^{(0)}_{\alpha}
-(-1)^{\alpha}\frac{Q}{2m}\right)\right].  \label{eq:2.37}
\end{equation}
Here, $Q=-n^{\tau}\int^{\infty}_{-{\infty}}dx^{\sigma}B(x)$; note that 
$Q$ in ${\sigma}_{\alpha}$ constitutes a Klein transformation.  We refer to
  ${\sigma}_{\alpha}$ as the spurion
 operator.
\cite{rf:13}  

We enumerate the properties of the ${\psi}_{\alpha}$ which show that the 
bosonized model is actually equivalent to the original model defined by the 
Lagrangian  (\ref{eq:2.1}).  We note that the symmetric energy-momentum tensor 
(\ref{eq:2.40})$\sim$(\ref{eq:2.43}) given below follows directly from the
 Lagrangian  (\ref{eq:2.10}).

(1) The Dirac equation is satisfied:
\begin{equation}
i{\gamma}^{\mu}({\partial}_{\mu}+ieA_{\mu}){\psi}=0  \label{eq:2.38}
\end{equation} 

(2) The canonical commutation relations with $A_{\sigma}$ and
 ${\pi}^{\sigma}$ 
and anticommutation relations are satisfied.

(3)  By applying the gauge invariant point-splitting procedure to e$\bar
{\psi}{\gamma}^{\mu}{\psi}$, we obtain the vector current
 $J^{\mu}=m{\partial}^
{\mu}X-m^2A^{\mu}=m{\varepsilon}^{{\mu}{\nu}}{\partial}_{\nu}{\lambda}$.
 This result
verifies that $j^{\mu}$ is given by $j^{\mu}=m{\partial}^{\mu}X$ so 
that it satisfies ${\varepsilon}^{{\mu}{\nu}}{\partial}_{\mu}j_{\nu}=0$.  The 
charge operator, $Q$, is given by
\begin{equation}
Q=\int^{\infty}_{-{\infty}}dx^{\sigma}J^{\tau}(x)=
-n^{\tau}\int^{\infty}_{-{\infty}}dx^{\sigma}B(x), 
  \label{eq:2.39}
\end{equation}
where the derivative terms integrate to zero.

(4) Applying the gauge invariant point-splitting procedure to the fermi 
products in the symmetric energy-momentum tensor and subtracting a divergent 
c-number~( we denote this procedure\cite{rf:14} by $R$), we get
\begin{eqnarray}
{\Theta}_{\tau}^{\;\;\sigma}&=&iR(\bar{\psi}{\gamma}^{\sigma}{\partial}_{\tau}
{\psi})-A_{\tau}J^{\sigma}-n^{\sigma}A_{\tau}B 
={\partial}_{\tau}{\lambda}{\partial}^{\sigma}{\lambda}
-n^{\sigma}A_{\tau}B, \\ \label{eq:2.40}
{\Theta}_{\tau}^{\;\;\tau}&=&-iR(\bar{\psi}{\gamma}^{\sigma}
{\partial}_{\sigma}{\psi})+A_{\sigma}J^{\sigma}
 +\frac{1}{2}(F_{\tau\sigma})^2 
-n^{\tau}BA_{\tau}
\nonumber \\
&=&-\frac{\cos 2\phi}{2}\biggl\{({\partial}_{\tau}{\lambda})^2+
({\partial}_{\sigma}{\lambda})^2 \biggr\}
+\frac{1}{2}(F_{\tau\sigma})^2-n^{\tau}BA_{\tau}, \\ \label{eq:2.41}
{\Theta}_{\sigma}^{\;\;\sigma}&=&iR(\bar{\psi}{\gamma}^{\sigma}
{\partial}_{\sigma}{\psi})-A_{\sigma}J^{\sigma}
 +\frac{1}{2}(F_{\tau\sigma})^2 
-n^{\sigma}BA_{\sigma} \nonumber \\
&=&\frac{\cos 2\phi}{2}\{({\partial}_{\tau}{\lambda})^2+({\partial}_{\sigma}
{\lambda})^2 \}
+\frac{1}{2}(F_{\tau\sigma})^2-n^{\sigma}BA_{\sigma}, \\ \label{eq:2.42}
{\Theta}_{\sigma}^{\;\;\tau}&=&iR(\bar{\psi}{\gamma}^{\tau}{\partial}_{\sigma}
{\psi}) -A_{\sigma}J^{\tau}-n^{\tau}BA_{\sigma} 
={\partial}_{\sigma}{\lambda}{\partial}^{\tau}{\lambda}
-n^{\tau}BA_{\sigma}.   \label{eq:2.43}
\end{eqnarray}

(5) Translational generators consist of those of the constituent fields: 
\begin{eqnarray}
P_{\tau}&=&\int^{\infty}_{-{\infty}}dx^{\sigma}:{\Theta}_{\tau}^{\;\;\tau}:
=\int^{\infty}_{-{\infty}}dx^{\sigma}:\biggl[-\frac{\cos 2\phi}{2}
\left\{ ({\partial}_{\tau}\tilde{\Sigma})^2+(
{\partial}_{\sigma}\tilde{\Sigma})^2\right\}+\frac{m^2}{2}(\tilde{\Sigma})^2
\nonumber \\&{}&\hspace{2.5cm}+\frac{1}{2}B\frac{n^2}
{m^2-n^2{\partial}_+^{\;2}}B+B\frac{n^{\sigma}}
{m^2-n^2{\partial}_+^{\;2}}{\partial}_{\sigma}C\biggr]:, \nonumber \\
P_{\sigma}&=&\int^{\infty}_{-{\infty}}dx^{\sigma}:{\Theta}_{\sigma}^{\;\;\tau}:
=\int^{\infty}_{-{\infty}}dx^{\sigma}:\left\{
 {\partial}_{\sigma}\tilde{\Sigma}
{\partial}^{\tau}\tilde{\Sigma}-B\frac{n^{\tau}}{m^2-n^2{\partial}_+^{\;2}}
{\partial}_{\sigma}C \right\}:.  \label{eq:2.44}
\end{eqnarray}

\section{Cancellation of Infrared divergences resulting from 
${\partial}_-^{-1}$}

We begin this section by pointing out that $X$ incorporates Higgs phenomena 
and that the possible infrared singularities in $A_{\mu}$ are in $X$.  
Therefore we confine ourselves to showing that $X$ is free from infrared 
divergences.  More precisely, we show that commutator function and the 
propagator for $X$ are free from infrared divergences.  To this aim we 
represent $\tilde{\Sigma}$ as 
\begin{equation}
\tilde{\Sigma}(x)=\frac{1}{\sqrt{4{\pi}}}\int^{\infty}_{-{\infty}}
\frac{dp_{\sigma}}{\sqrt{p^{\tau}}}
\{ a(p_{\sigma})e^{-ip{\cdot}x}+a^{\ast}(p_{\sigma})e^{ip{\cdot}x} \}.
\label{eq:3.1}
\end{equation}
Here, $p^{\tau}=\sqrt{p_{\sigma}^{\;2}+m_0^{\;2}}$ with $m_0^{\;2}
=-\cos 2\phi m^2$ 
and 
\begin{equation}
[a(p_{\sigma}),a(q_{\sigma})]=0, \quad 
[a(p_{\sigma}),a^{\ast}(q_{\sigma})]={\delta}(p_{\sigma}-q_{\sigma}).
\end{equation}

We first show that the commutator function of $X$ includes the commutator 
function, $E(x)$, characteristic of a dipole ghost field.   From 
(\ref{eq:2.28}), (\ref{eq:2.32}) and (\ref{eq:3.1}) we obtain
\begin{equation}
[X(x),X(y)]=i\{{\Delta}(x-y;m^2)+n^2m^2E(x-y)\} \label{eq:3.3}
\end{equation}
where ${\Delta}(x;m^2)$ is the commutator function of the free boson field 
of mass $m$ and 
\begin{eqnarray}
E(x)&=&\frac{1}{{\partial}_-^{\;2}}{\Delta}(x;m^2)
 -\frac{x^{\tau}}{m^2(n^{\tau})^2-n^2{\partial}_{\sigma}^{\;2}}
{\delta}(x^{\sigma}-\cot(\theta-\phi)x^{\tau}) \nonumber \\
&+&\frac{2n^{\tau}n_{\sigma}}{(m^2(n^{\tau})^2-n^2{\partial}_{\sigma}^{\;2})^2}
{\partial}_{\sigma}{\delta}(x^{\sigma}-\cot(\theta-\phi)x^{\tau}).
\label{eq:3.4}
\end{eqnarray}
When $x^{\tau}=y^{\tau}$, the commutator $[X(x),X(y)]$ vanishes.   We see that 
as follows: The first term of (\ref{eq:3.3}) vanishes trivially.  The second 
term of $E(x-y)$, which is proportional to $x^{\tau}-y^{\tau}$, also vanishes 
trivially.  If we evaluate the first term  of (\ref{eq:3.4}) using
${\partial}_-^{-1}$ as defined in (\ref{eq:2.26}), then we get a nonvanishing 
term;  however, that term is cancelled by the third term of (\ref{eq:3.4}).  

The following are properties of $E(x)$:
\begin{eqnarray}
&({\square} +m^2)E(x)=-\frac{x^{\tau}}{(n^{\tau})^2}{\delta}(x^{\sigma}
-\cot(\theta-\phi)x^{\tau}), \quad {\partial}_-^{\;2}E(x)={\Delta}(x;m^2),&
 \\
&E(x)|_{x^{\tau}=0}=0, \quad {\partial}_-E(x)|_{x^{\tau}=0}=0, \quad 
{\partial}_-^{\;2}E(x)|_{x^{\tau}=0}=0, & \\
&({\square} +m^2)E(x)|_{x^{\tau}=0}=0, \quad 
(\square +m^2){\partial}_-E(x)|_{x^{\tau}=0}=-\frac{1}{n^{\tau}}{\delta}
(x^{\sigma}).& \label{eq:3.7}
\end{eqnarray}

Next we show that the vacuum expectation value, ${\langle}0|X(x)X(y)|0
{\rangle}$, does not diverge when $x^{\tau}=y^{\tau}$.  We will need to use 
${\langle}0|X(x)X(y)|0{\rangle}$ evaluated at $x^{\tau}=y^{\tau}$ to calculate 
the equal $x^{\tau}$-time anticommutatuion relations of the fermion field 
operators.  It is straightforward to obtain
\begin{equation}
{\langle}0|X(x)X(y)|0{\rangle}={\Delta}^{(+)}(x-y;m^2)+m^2n^2E^{(+)}(x-y)
 \label{eq:3.8}
\end{equation}
where ${\Delta}^{(+)}(x;m^2)$ is the positive frequency part of $i{\Delta}
(x;m^2)$ and 
\begin{eqnarray}
E^{(+)}(x)&=&\frac{1}{{\partial}_-^{\;2}}{\Delta}^{(+)}(x;m^2)
 -\frac{ix^{\tau}}{2\pi}\int_{-\infty}^0dk_{\sigma}
\frac{1}{m^2(n^{\tau})^2+n^2k_{\sigma}^{\;2}}{\rm e}^{-ik{\cdot}x} \nonumber
 \\
&+&\frac{1}{2\pi}\int_{-\infty}^0dk_{\sigma}\frac{2n^{\tau}n_{\sigma}
k_{\sigma}}{(m^2(n^{\tau})^2+n^2k_{\sigma}^{\;2})^2}{\rm e}^{-ik{\cdot}x}.
\label{eq:3.9}
\end{eqnarray}
A logarithmic divergence appears in the second term 
but we regularize it with the principal value prescription.  In addition,
 linear divergences appear in the first and third terms when $x^{\tau}=0$.  
 We set $x^{\tau}=0$ and divide the integration region of the first term into 
a region where the integration variable is positive and a region where the 
integration variable is negative.   We then combine the integration in the 
negative region with third term and obtain  
\begin{eqnarray}
\hspace*{-7mm}&{}&\!\!\frac{-1}{4\pi}\!\!\int_{-\infty}^0\frac{dp_{\sigma}}
{p^{\tau}}\frac{1}{p_-^{\;2}}
{\rm e}^{-ip_{\sigma}x^{\sigma}}+\frac{1}{2\pi}\int_{-\infty}^0dk_{\sigma}
\frac{2n^{\tau}n_{\sigma}k_{\sigma}}{(m^2(n^{\tau})^2+n^2k_{\sigma}^{\;2})^2}
{\rm e}^{-ik_{\sigma}x^{\sigma}} \nonumber \\
\hspace*{-7mm}&=&\hspace*{-1.5mm}\frac{-1}{4\pi}\!\!\int_{-\infty}^0\!\!
\frac{dp_{\sigma}}{p^{\tau}}
\frac{(n^{\tau}p^{\tau}-n_{\sigma}p_{\sigma})^2}
{(m^2(n^{\tau})^2+n^2p_{\sigma}^{\;2})^2}
{\rm e}^{-ip_{\sigma}x^{\sigma}}
\hspace*{-1.5mm}=\hspace*{-1.5mm}\frac{-1}{4\pi}\!\!
\int_{-\infty}^0\!\! \frac{dp_{\sigma}}{p^{\tau}}
\frac{(-\cos 2\phi)^2}{(n^{\tau}p^{\tau}+n_{\sigma}p_{\sigma})^2}
{\rm e}^{-ip_{\sigma}x^{\sigma}}\!\!.  \label{eq:3.10}
\end{eqnarray}
It is useful to recall here that $\phi$ and $\theta$ lie in the regions 
 $(\frac{\pi}{4}<\theta<\phi{\leqq}\frac{\pi}{2})$ so that 
$n^{\tau}=\sin(\phi-\theta)>0$ and $n_{\sigma}
=\cos(\phi+\theta)<0$.  As a result, no infrared divergences appear from 
(\ref{eq:3.10}).  Furthermore, changing the integration variable from 
$p_{\sigma}$ to $-p_{\sigma}$ verifies that (\ref{eq:3.10}) is equal to the 
positive integration part of the first term of $E^{(+)}(x)$.  It follows that 
$E^{(+)}(x-y)$ is well defined at $x^{\tau}=y^{\tau}$, which implies that 
we can incorporate  the equal $x^{\tau}$-time anticommutation relations of the 
fermion field operators in the same way as \S~3 in I.

Finally, we show that the factors $(m^2(n^{\tau})^2+n^2p_{\sigma}^{\;2})
^{-1}$ relevant to the infrared divergences drop out completely from 
the propagator for $X$, which is given by 
\begin{equation}
{\langle}0|T(X(x)X(y))|0{\rangle}={\Delta}_{\scriptscriptstyle F}(x-y;m^2)
+m^2n^2E_{\scriptscriptstyle F}(x-y)
 \label{eq:3.11}
\end{equation}
where ${\Delta}_{\scriptscriptstyle F}(x-y;m^2)$ is the propagator for the 
free boson field of mass $m$ and $E_{\scriptscriptstyle F}(x-y)$ is defined
 by
\begin{eqnarray}
E_{\scriptscriptstyle F}(x-y)&=&\theta(x^{\tau}-y^{\tau})E^{(+)}(x-y)
+\theta(y^{\tau}-x^{\tau})E^{(+)}(y-x) \nonumber \\
&=&\frac{1}{(2\pi)^2}\int d^2q
E_{\scriptscriptstyle F}(q){\rm e}^{-iq{\cdot}(x-y)}.   \label{eq:3.12}
\end{eqnarray}
Substituting the expression in (\ref{eq:3.9}) into (\ref{eq:3.12}) and 
then Fourier transforming it provides us with
\begin{eqnarray}
\hspace*{-1cm}&{}&E_{\scriptscriptstyle
 F}(q)=-\frac{i}{q^2-m^2+i{\varepsilon}}
\frac{(n^{\tau})^2(q_{\sigma}^{\;2}-\cos2\phi m^2)+n_{\sigma}^{\;2}
q_{\sigma}^{\;2}+2n^{\tau}n_{\sigma}q^{\tau}q_{\sigma}}
{(m^2(n^{\tau})^2+n^2q_{\sigma}^{\;2})^2} \nonumber \\
\hspace*{-1cm}&+&\frac{i}{(q_-+i{\varepsilon}{\rm sgn}(q_+))^2 }
\frac{(n^{\tau})^2}
{m^2(n^{\tau})^2+n^2q_{\sigma}^{\;2}}+\frac{i}{q_-+i{\varepsilon}
{\rm sgn}(q_+)}\frac{2(n^{\tau})^2n_{\sigma}q_{\sigma}}
{(m^2(n^{\tau})^2+n^2q_{\sigma}^{\;2})^2}  
\end{eqnarray}
where $q_+{\equiv}-\frac{q_{\sigma}}{n^{\tau}}$.   The term on the 
first line comes from the physical degrees of freedom whereas the 
terms on the second line come from the residual gauge degrees of freedom.  
It is remarkable that if we combine them into one term, then we obtain
$E_{\scriptscriptstyle F}(q)=-\frac{i}{q^2-m^2+i{\epsilon}}{\times}
\frac{1}{(q_-+i{\varepsilon}{\rm sgn}(q_+))^2 }$ and hence
\begin{equation}
{\langle}0|T(X(x)X(y))|0{\rangle}\!=\!\frac{1}{(2\pi)^2}\!\!\int d^2q
\frac{i}{q^2-m^2+i{\epsilon}}\left(1-\frac{n^2m^2}{(q_-+i{\varepsilon}
{\rm sgn}(q_+))^2 }\right){\rm e}^{-iq{\cdot}(x-y)}.  \label{eq:3.14} 
\end{equation}
In this way, the infrared divergences are eliminated and the singularity 
associated with the gauge fixing is prescribed in such a way that 
causality is preserved in complex $q_{\tau}$ coordinates.  It can be shown
 that the same is true of the propagator for $A_{\mu}$ and we get 
\begin{equation}
\int d^2x{\langle}0|T(A_{\mu}(x)A_{\nu}(0))|0{\rangle}
{\rm e}^{iq{\cdot}x}=\frac{iP_{{\mu}{\nu}}}{q^2-m^2+i{\epsilon}} 
\end{equation}
where  
\begin{equation}
P_{{\mu}{\nu}}= -g_{{\mu}{\nu}}
+\frac{n_{\mu}q_{\nu}+n_{\nu}q_{\mu}}{q_-+i{\varepsilon}{\rm sgn}(q_+)}-
n^2\frac{q_{\mu}q_{\nu}}{(q_-+i{\varepsilon}{\rm sgn}(q_+))^2 }.  
 \label{eq:3.16} 
\end{equation}

\section{Pure space-like case }

We begin by noting that the limit $\phi{\to}\theta$ of the residual gauge part 
of the operator solution given in \S~2, which has factors dependent on 
quantization coordinates, is not well-defined.   We see from this that 
an operator solution in the (PSL) case, $\phi=\theta$, 
is not constructed in the same manner as that given in \S~2.  

The Lagrangian and the equations of motion for $A_{\mu}$ and $X$ in the PSL 
case are given respectively by transforming (\ref{eq:2.1}),(\ref{eq:2.7}) and 
(\ref{eq:2.8}) into those in $+-$-coordinates.  We obtain two new constraints 
\begin{equation}
{\pi}^-+{\partial}_-A_+=0, \quad 
{\partial}_-{\pi}^--m{\pi}_{\scriptscriptstyle X}=0 \label{eq:4.1}
\end{equation}
in addition to the gauge fixing condition $A_-=0$.
As a result only $X$ and ${\pi}_{\scriptscriptstyle X}$ are 
left as independent canonical variables.  This reflects the fact that the 
residual gauge fields depend on only $x^+$ so they cannot 
be canonical variables.  Therefore we cannot obtain 
their quantization conditions from the Dirac procedure.\cite{rf:15} Instead
 we employ the following items as guiding principles to introduce them in the
 PSL case:\\
(1) $X$ and ${\pi}_{\scriptscriptstyle X}$ satisfy the canonical 
commutations conditions.  \\
(2) The residual gauge fields commute with the massive field.\\
(3) $B$ satisfies $[B(x),X(y)]=im\delta(x^+-y^+)$ and so, generates 
$c$-number residual gauge transformations.\\
(4) The infrared divergences which come from the physical part of $X$ 
are regularized by infrared divergences from the residual gauge fields.

We start constructing an operator solution by solving Eq.(2.8) and obtain 
an expression similar to (2.17).  The massive field obtained will be identified
 below as $\tilde{\Sigma}$.  So, since $A_-  = 0$, we can write 
\begin{equation}
{\partial}_-X={\partial}^+\tilde{\Sigma}-\frac{mn^2}{m^2-n^2{\partial}_+^{\;2}}
B.  \label{eq:4.2}
\end{equation}
We see from this that $X$ is given by
$$X=\frac{{\partial}^+}{{\partial}_-}\tilde{\Sigma}
-\frac{mn^2x^-}{m^2-n^2{\partial}_+^{\;2}}B +{\rm integration~ constant}.$$ 
The massive, physical part of $X$ is now known and 
${\pi}_{\scriptscriptstyle X}$ can be written as
\begin{equation}
{\pi}_{\scriptscriptstyle X}={\partial}^+X-mA^+
={\partial}_-{\lambda}={\partial}_-\tilde{\Sigma}, \label{eq:4.3}
\end{equation}
>From this we see that if we impose the canonical commutation conditions on $X$ 
and ${\pi}_{\scriptscriptstyle X}$, that will imply the following equal $x^+$ -time commutation relations 
\begin{equation}
[\tilde{\Sigma}(x),\tilde{\Sigma}(y)]=0, \;\; 
[\tilde{\Sigma}(x),{\partial}^+\tilde{\Sigma}(y)]=i\delta(x^--y^-), \;\;
[{\partial}^+\tilde{\Sigma}(x),{\partial}^+\tilde{\Sigma}(y)]=0.  \label{eq:4.4}
\end{equation}
The integration constant is determined in the following way: To implement 
the residual gauge transformation, we add
 $\frac{m}{m^2-n^2{\partial}_+^{\;2}}
C$ to $X$, where $C$ is the conjugate to $B$ and satisfies the following 
commutation relations
\begin{equation}
[C(x),C(y)]=0,\;\;[B(x),\frac{1}{m^2-n^2{\partial}_+^{\;2}}C(y)]
=i{\delta}(x^+-y^+).  
\label{eq:4.5}
\end{equation} 
Furthermore we require that the infrared divergence resulting from 
$\frac{{\partial}^+}{{\partial}_-}\tilde{\Sigma}$ be cancelled through the 
mechanism worked out in \S~2.  If we consider a surface of constant $x^-$, then 
we can write  
\begin{equation}
\frac{{\partial}^+}{{\partial}_-}\tilde{\Sigma}
=\frac{n_+m^2+n_-{\partial}_+{\partial}^-}
{m^2-n^2{\partial}_+^{\;2}}\tilde{\Sigma} \label{eq:4.6}
\end{equation}
and see that the equal $x^-$-time commutator, 
$[\frac{{\partial}^+}{{\partial}_-}\tilde{\Sigma}(x),\frac{{\partial}^+}
{{\partial}_-}\tilde{\Sigma}(y)]$, does not vanish.
To correct for this we must add 
$\frac{mn_+n_-}{(m^2-n^2{\partial}_+^{\;2})^2}{\partial}_+B$ to 
$X$.  Summing all the terms, we obtain
\begin{equation}
X=\frac{{\partial}^+}{{\partial}_-}\tilde{\Sigma}
 +\frac{m}{m^2-n^2{\partial}_+^{\;2}}\left(C-n^2x^-B
+\frac{mn_+n_-}{m^2-n^2{\partial}_+^{\;2}}{\partial}_+B\right), 
\label{eq:4.7}
\end{equation}
This is exactly the solution given in I.  Therefore, we need not repeat the 
construction of the fermion field operators or the description of their properties.  

It remains to be shown that infrared divergences do not appear 
when we evaluate $\langle0|X(x)X(y)|0\rangle$ at $x^+=y^+$ or when 
we calculate the $x^+$-time ordered propagator for $X$.   By 
using the representations for the constituent operators given 
in I, we obtain the following vacuum expectation value 
\begin{equation}
\langle0|X(x)X(y)|0\rangle={\Delta}^{(+)}(x-y;m^2)+n^2m^2E^{(+)}_{PSL}(x-y)
\label{eq:4.8}
\end{equation}
where
\begin{eqnarray}
&{}&E^{(+)}_{PSL}(x)=-\frac{1}{4\pi}\int_{-\infty}^{\infty}
\frac{dp_-}{p^+}\frac{{\rm e}^{-ip{\cdot}x}}{p_-^{\;2}} \nonumber \\
&-&\frac{ix^-}{2\pi}\int_0^{\infty}dk_+\frac{{\rm e}^{-ik_+x^+}}
{m^2+n^2k_+^{\;2}}+\frac{1}{2\pi}\int_0^{\infty}dk_+
\frac{2n_+k_+{\rm e}^{-ik_+x^+}}{(m^2+n^2k_+^{\;2})^2}.  \label{eq:4.9}
\end{eqnarray}
Here, $p^+$ and $p_+$ are defined, respectively, as
 $p^+=\sqrt{p_-^{\;2}+m^2_0},\;
(m_0^2=-n^2m^2),\;p_+=\frac{p^+-n_+p_-}{-n_-}$.
The integral on the first line comes from the $\tilde{\Sigma}$, while 
the integrals  on the second line come from the residual gauge fields.  
The value of $E^{(+)}_{PSL}(x)$ at $x^+=0$ is formally given by
\begin{eqnarray}
\hspace*{-8mm}&{}&E^{(+)}_{PSL}(x)|_{x^+=0}=-\frac{1}{4\pi}
\int_{-\infty}^{\infty}
\frac{dp_-}{p^+}\frac{{\rm e}^{-ip_-x^-}-1}{p_-^{\;2}} \nonumber \\
\hspace*{-8mm}&-&\frac{1}{4\pi}\int_{-\infty}^{\infty}
\frac{dp_-}{p^+}\frac{1}{p_-^{\;2}}
-\frac{ix^-}{2\pi}\int_0^{\infty}dk_+\frac{1}
{m^2+n^2k_+^{\;2}}+\frac{1}{2\pi}\int_0^{\infty}dk_+
\frac{2n_+k_+}{(m^2+n^2k_+^{\;2})^2} \label{eq:4.10}
\end{eqnarray}
where we have divided the first term into a finite term and a diverging term.  
It should be noted here that $p_-$ is conjugate to 
the spatial variable $x^-$, while $k_+$ is conjugate to the temporal
 variable $x^+$.  To make the infrared divergence cancellation 
mechanism work as in \S~3, both integration variables have to be spatial or 
temporal.  Therefore, we change the integration variable from the spatial 
$p_-$ to the temporal $p_+=\frac{\sqrt{p_-^{\;2}+m_0^{\;2}}-n_+p_-}{-n_-}$.  
At the same time we denote $k_+$ as $p_+$.  If we take account of the fact
 that $p_+$ is two-valued function of $p_-$, then we can rewrite the diverging 
integral into the following form
\begin{eqnarray}
\int_{-\infty}^{\infty}\frac{dp_-}{p^+}\frac{1}{p_-^{\;2}}
&=&\int_{m_0}^{\infty}\frac{dp_+}{\sqrt{p_+^{\;2}-m_0^{\;2}}}
\left(\frac{-n_-}{n_+p_+-\sqrt{p_+^{\;2}-m_0^{\;2}}}\right)^2 \nonumber \\
&+&\int_{m_0}^{\infty}\frac{dp_+}{\sqrt{p_+^{\;2}-m_0^{\;2}}}
\left(\frac{-n_-}{n_+p_++\sqrt{p_+^{\;2}-m_0^{\;2}}}\right)^2
 \label{eq:4.11}
\end{eqnarray} 
where the first term diverges, but second term is finite.  Now we see 
that if we combine the first integral in (\ref{eq:4.11}) with the third one 
on the second line of (\ref{eq:4.10}), we obtain the following finite integrals:
\begin{eqnarray}
&-&\int_{m_0}^{\infty}\frac{dp_+}{\sqrt{p_+^{\;2}-m_0^{\;2}}}
\left(\frac{-n_-}{n_+p_+-\sqrt{p_+^{\;2}-m_0^{\;2}}}\right)^2
+\int_0^{\infty}dp_+\frac{4n_+p_+}{(m^2+n^2p_+^{\;2})^2} \nonumber \\
&=&-\int_{m_0}^{\infty}\!\frac{dp_+}{\sqrt{p_+^{\;2}-m_0^{\;2}}}
\left(\frac{n_+p_++\sqrt{p_+^{\;2}-m_0^{\;2}}}{m^2+n^2p_+^{\;2}}\right)^2
+\int_0^{\infty}dp_+\frac{4n_+p_+}{(m^2+n^2p_+^{\;2})^2} \nonumber \\
&=&-\int_{m_0}^{\infty}\frac{dp_+}{\sqrt{p_+^{\;2}-m_0^{\;2}}}
\left(\frac{n_+p_+-\sqrt{p_+^{\;2}-m_0^{\;2}}}
{m^2+n^2p_+^{\;2}}\right)^2
+\int_0^{m_0}dp_+\frac{4n_+p_+}{(m^2+n^2p_+^{\;2})^2} \nonumber \\
&=&\!-\!\int_{m_0}^{\infty}\!\frac{dp_+}{\sqrt{p_+^{\;2}-m_0^{\;2}}}\!
\left(\frac{-n_-}{n_+p_++\sqrt{p_+^{\;2}-m_0^{\;2}}} \right)^2
\!+\!\int_0^{m_0}dp_+\frac{4n_+p_+}{(m^2+n^2p_+^{\;2})^2}.  \label{eq:4.12}
\end{eqnarray}
After tedious but straightforward calculations, we finally obtain
\begin{equation}
E^{(+)}_{PSL}(x)|_{x^+=0}=\frac{1}{4\pi}
\int_{-\infty}^{\infty}
\frac{dp_-}{p^+}\frac{1-\cos(p_-x^-)}{p_-^{\;2}}
+\frac{1}{2\pi m^2}\frac{-n_-}{1+n_+}.  \label{eq:4.13}
\end{equation}

Finally, without demonstration, we give the $x^+$-time ordered propagator for 
$X$.  The necessary demonstration can be carried out in parallel with that given in Appendix A in I.  It turns out that 
\begin{equation}
\langle0|T(X(x)X(y))|0\rangle\!=\!\frac{1}{(2\pi)^2}
\!\int\! d^2q\frac{i}{q^2-m^2+i{\varepsilon}}\!\left(1-\frac{n^2m^2}
{(q_-+i{\varepsilon}{\rm sgn}(q_+))^2}\right){\rm e}^{-iq{\cdot}(x-y)}.
\label{eq:4.14}
\end{equation} 
It is remarkable to see that in spite of the fact that all factors depending 
on the quantization coordinates drop out, we have the same 
propagator that we have obtained in (\ref{eq:3.14}).  

\section{Concluding remarks}

In this paper the framework used in I has been generalized by introducing 
the $\tau\sigma$-coordinates and at the same time simplified by bosonizing
 the model.  The new framework has allowed us to investigate the way in which 
 operator solutions develop a dependence on the quantization coordinates.  
 In the new framework we can take the dipole ghost field, $X$, and the 
 component of the gauge field, $A_{\sigma}$, as canonical variables.  We have 
given special attention to the determination of  $X$, because we know that it 
cannot be a manifest Lorentz scalar since it develops an explicit dependence on 
the quantization coordinates.  We have found that the physical part of $X$ is 
determined uniquely by the gauge choice, while the residual gauge part, which 
contains the manifest dependence on the quantization coordinates, is determined 
by requiring that $X$ satisfy the canonical commutation conditions.  The main 
findings of this paper are:  \\
(1)the residual gauge fields are indispensable ingredients of the space-like 
axial gauge Schwinger model.\\
(2) $(n{\cdot}{\partial})^{-1}=({\partial}_-)^{-1}$ is ill-defined 
irrespective of the quantization coordinates, as long as the gauge fixing 
direction is space-like ($n^2<0$).  \\
(3) As a consequence, the physical part of $X$ includes infrared 
divergences irrespective of the quantization coordinates, as long as the gauge 
fixing direction is space-like.\\
(4) If we introduce the residual gauge fields in such a way that $X$ 
satisfies the canonical commutation relations, then the residual gauge
 part is determined so as to regularize the infrared divergences resulting 
from the physical part.  

In the PSL case the residual gauge fields cannot be canonical variables due
 to the fact that they depend on the evolution parameter $x^+$. So the 
operator solution for this case cannot be constructed purely by canonical 
methods.  We have overcome this difficulty by employing the items described in 
\S~4 as guiding principles supplement canonical methods. The operator solution 
we obtain by the extended methodology is satisfactory in every aspect.  In 
particular, all ill-defined factors drop out from the $x^+$-time ordered 
propagators for $X$ and $A^{\mu}$ so that we have the same ML form for the 
propagators irrespective of the quantization coordinates.  

The light-cone gauge, $\theta=\frac{\pi}{4}$, is exceptional.  In 
that case, $n^2$ is zero, so the manifest dependence on the quantization 
coordinates disappears whatever coordinates we may have.  Therefore, 
the light-cone axial gauge formulation is obtained  by simply setting 
${\theta}=\phi=\frac{\pi}{4}$ and we obtain
\begin{equation}
X=\tilde{\Sigma}+m^{-1}C, \quad A_+=2\frac{{\partial}_+}{m}\tilde{\Sigma}
+\frac{{\partial}_+}{m^2}(C+{\partial}_+^{-1}B).
\end{equation} 
On comparing these operators with the corresponding operators we gave in a 
previous  paper,\cite{rf:16} we find that $B$ and $C$ are related to $\eta$ 
and $\phi$ by $C=m(\eta+\phi), \; B=m{\partial}_+(\eta-\phi).$

We end this paper by pointing out that in axial gauge quantizations in 
4-dimensions, the same infrared divergence cancellation mechanism works.  We 
already have some work in this direction.\cite{rf:17}  We hope to report more 
in subsequent studies.


\begin{thebibliography}{99}
\bibitem{rf:1}
Y.~Nakawaki and G.~McCartor, \PTP{106,2001,167}.  

\bibitem{rf:2} This procedure was first devised in G.~McCartor, 
Z.~Phys.~\andvol{C41,1988, 271}.

\bibitem{rf:3}
N.~Nakanishi, \JL{Phys.~Lett.,131B,1983, 381}.    
N.~Nakanishi, {\it Quantum Electrodynamics},~ed.T.~Kinoshita 
(World Scientific, Singapore, 1990), p.~36.

\bibitem{rf:4}
N.~Nakanishi, \PTP{67,1982,965}.  \\ 
K.~Haller, \JL{Phys.~Lett.,251B,1990, 575}.\\
P.~V.~Landshoff and P.~van ~Nieuwenhuizen, \PR{D50,1994,4157}.  \\ 
M.~Morara and R.~Soldati, \PR{D58,1998,105011}.    

\bibitem{rf:5}
G.~McCartor and D.G.~Robertson, 
Z.~Phys.~\andvol{C62,1994, 349}; \andvol{C68,1995,345}.

\bibitem{rf:6}
S.~Mandelstam, Nucl.~Phys.\ {\bf B213} (1983), 149.  

\bibitem{rf:7}
G.~Leibbrandt, Phys.~Rev.\ {\bf D29} (1984), 1699.    

\bibitem{rf:8}
A.~Bassetto, ~M.~Dalbosco,~I.~Lazziera and R.~Soldati, 
        Phys.~Rev.\ {\bf D31} (1985), 2012.    
\\
A.~Bassetto,~G.~Nardelli and R.~Soldati, {\it Yang-Mills Theories in
 Algebraic
 Non-Covariant Gauges} (World Scientific, Singapore, 1991).

\bibitem{rf:9}
I.~Lazzizzera, \PL{210B,1988,188}; Nuovo Cim.  \andvol{102A,1989,1385}.  

\bibitem{rf:10}
B.~Lautrup, Mat.~Fys.~Medd.~Dan.~Vid.~Selsk.~{\bf 35}(1967),~No.11.   \\
N.~Nakanishi, Prog.  Theor.  Phys.  Suppl.  No.51~(1972),1.

\bibitem{rf:11} 
J.~H.~Lowenstein and J.~A.~Swieca, Ann.  of Phys.~\andvol{68,1971,172}.\\
N.~Nakanishi, \PTP{57,1977,580}.

\bibitem{rf:12}
J.~Schwinger, \PRL{3,1959,296}.  

\bibitem{rf:13}
Y.~Nakawaki, \PTP{72,1984,134}.

\bibitem{rf:14}
Y.~Nakawaki, \PTP{64,1980,1828}.

\bibitem{rf:15}
P.~A.~M.~Dirac, {\it Lectures on Quantum Mechanics}, Belfer Graduate School
 of 
\\
Science-Yeshiva University (Academic Press, New York, 1964).
\\
A.~Hanson,~T.~Regge and C.~Teitelboim, {\it Constrained Hamiltonian Systems} 
\\(Accad.~Naz.~deiLincei, Rome, 1976).

\bibitem{rf:16}
Y.~Nakawaki and G.~McCartor, \PTP{103,2000,161}.  

\bibitem{rf:17}
Y.~Nakawaki and G.~McCartor, \PTP{103,1999,149}; 
\NP{B108(proc.Suppl.),2002,201}.  


\end{thebibliography}
\end{document}